# Subunit rotation in a single $F_oF_1$-ATP synthase in a living bacterium monitored by FRET

K. Seyfert[#a], T. Oosaka[#b], H. Yaginuma[b], S. Ernst[a], H. Noji[b, c], R. Iino*[b], M. Börsch*[a]

[a]Third Institute of Physics, University of Stuttgart, Pfaffenwaldring 57, 70550 Stuttgart, Germany
[b]Institute of Scientific and Industrial Research, Osaka University, Ibaraki 567-0047, Osaka, Japan
[c]Department of Applied Chemistry, The University of Tokyo, Hongo 7-3-1, Tokyo, Japan

**ABSTRACT**

$F_oF_1$-ATP synthase is the ubiquitous membrane-bound enzyme in mitochondria, chloroplasts and bacteria which provides the 'chemical energy currency' adenosine triphosphate (ATP) for cellular processes. In *Escherichia coli* ATP synthesis is driven by a proton motive force (PMF) comprising a proton concentration difference $\Delta pH$ plus an electric potential $\Delta\Psi$ across the lipid membrane. Single-molecule *in vitro* experiments have confirmed that proton-driven subunit rotation within $F_oF_1$-ATP synthase is associated with ATP synthesis. Based on intramolecular distance measurements by single-molecule fluorescence resonance energy transfer (FRET) the kinetics of subunit rotation and the step sizes of the different rotor parts have been unraveled. However, these experiments were accomplished in the presence of a PMF consisting of a maximum $\Delta pH \sim 4$ and an unknown $\Delta\Psi$. In contrast, in living bacteria the maximum $\Delta pH$ across the plasma membrane is likely 0.75, and $\Delta\Psi$ has been measured between -80 and -140 mV. Thus the problem of *in vivo* catalytic turnover rates, or the *in vivo* rotational speed in single $F_oF_1$-ATP synthases, respectively, has to be solved. In addition, the absolute number of functional enzymes in a single bacterium required to maintain the high ATP levels has to be determined. We report our progress of measuring subunit rotation in single $F_oF_1$-ATP synthases *in vitro* and *in vivo,* which was enabled by a new labeling approach for single-molecule FRET measurements.

**Keywords:** Rotary motor, $F_oF_1$-ATP synthase, single-molecule detection, Förster resonance energy transfer FRET, FRET imaging

## 1 INTRODUCTION

Adenosine triphosphate (ATP) is one of the most important compounds for living organisms. Cells consume ATP for their varieties of functions to support their life. The ATP-producing biological nanomachine $F_oF_1$-ATP synthase ($F_oF_1$) is working in all types of cells, such as bacteria, plants and animals.

$F_oF_1$, a membrane-embedded enzyme, consists of two distinct parts which are operating as two coupled rotary motors. In the plasma membranes of *Eschericha coli* (*E. coli*) bacteria, the proton-driven $F_o$ motor is powering the ATP-synthesizing $F_1$ part. The proton transport in the $F_o$ motor is tightly coupled with ATP synthesis and hydrolysis in the $F_1$ motor, and the direction of the reaction is determined by the balance between the amplitude of the proton motive force (PMF) and the free energy of ATP hydrolysis that depends on the concentrations of ATP, ADP and inorganic phosphate in the cell.

The "rotary catalysis" of $F_oF_1$ was first proposed by Boyer based on the elaborated biochemical studies[1-4], and strongly supported by Walker's first crystal structure of the isolated $F_1$ part[5]. Rotation of $F_1$ driven by ATP hydrolysis was

..................................................................................................................................................................

\#      Equal contributions

\*      iino@sanken.osaka-u.ac.jp; phone (81) 6 6879 8481; fax (81) 6 6875 5724

       m.boersch@physik.uni-stuttgart.de; phone (49) 711 6856 4632; fax (49) 711 6856 5281; http://www.m-boersch.org

further supported by fluorescence anisotropy measurements[6, 7], and was directly proven by the single-molecule observation with a large probe[8-12]. The ATP-driven subunit rotation in the whole complex $F_oF_1$ was also proved by single-molecule observation using a large probe[13-15].

In contrast to the rotation driven by ATP hydrolysis, rotation driven by the PMF was more difficult to prove because $F_oF_1$ had to be embedded in the membrane. This was first accomplished by the single-molecule Förster resonance energy transfer (FRET) measurement in which donor and acceptor dyes were attached to the rotor and stator subunits of $F_oF_1$ reconstituted in to the liposome[16-22]. The components of PMF, $\Delta$pH and $\Delta\Psi$, were applied by the acid-base transition and the $K^+$-valinomycin diffusion potential[23-25].

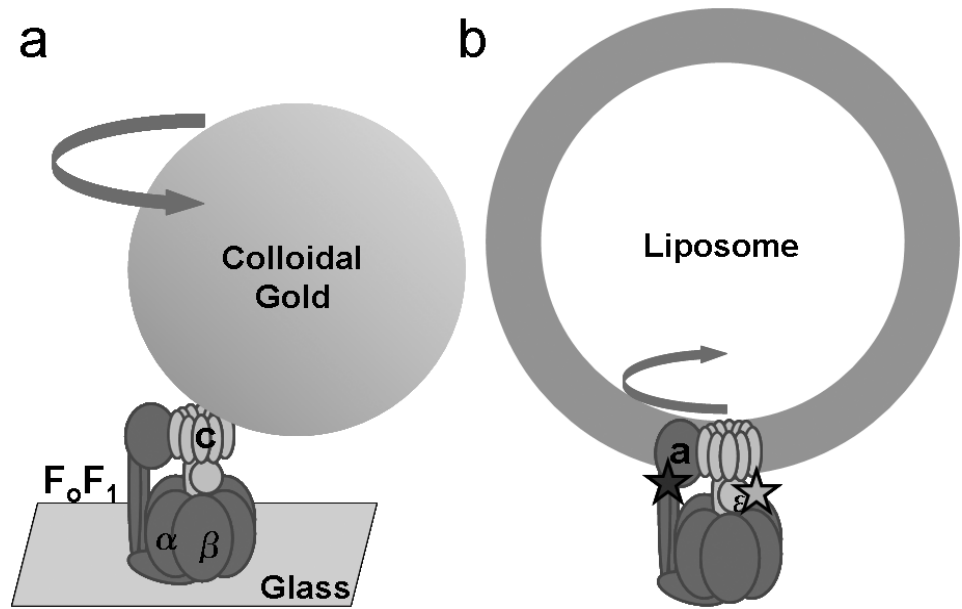

**Figure 1: Principles of single-molecule rotation measurements in $F_oF_1$-ATP synthase.** (**a**) Rotation monitored by a large light-scattering probe attached to the c-subunits[14]. The enzyme is attached to the glass surface *via* Histidine-tags, and subunit rotation is observed by video microscopy. (**b**) Rotation measurements by intramolecular Förster-type resonance energy transfer (FRET)[22]. The FRET-labeled enzyme (symbolized by the two stars on the rotating $\varepsilon$- and the static *a*-subunit) is reconstituted into a liposome and is freely diffusing in solution.

*E. coli* has two ATP synthesis pathways, glycolysis and oxidative phosphorylation. $F_oF_1$ is responsible for ATP synthesis in the later process. Although the rotation of $F_oF_1$ driven by PMF *in vitro* has been proven as described above, there are three fundamental questions on the ATP synthesis rate of $F_oF_1$ in the cell. 1) *E. coli* cells deficient in $F_oF_1$ can grow in normal culture condition in the lab. This suggests that ATP supply by glycolysis is sufficient. 2) In the cell, concentration of ATP is much higher than that of ADP. This shifts the equilibrium of the reaction catalyzed by $F_oF_1$ toward ATP hydrolysis. 3) The proton motive force (PMF), especially the $\Delta$pH, that drives the ATP synthesis of $F_oF_1$, is not so large in *E. coli*. In this study, using the single-molecule FRET technique, we aim at measuring the ATP synthesis rate of $F_oF_1$ working in living cells. For single-molecule FRET measurements *in vivo*, the rotor and stator of $F_oF_1$ in the cell have to be labeled by donor and acceptor dyes efficiently and specifically, and the copy numbers of labeled $F_oF_1$ have to be lowered to the density at which single molecules can be distinguished in the cell.

The former task was accomplished by introduction of SNAP- and CLIP-tags[26] or Tetracysteine-tag[27] to the $F_oF_1$, and by specific labeling of of the tags with benzylguanine-, benzylcytosine- and biarsenical-derivatives of fluorescent dyes. The SNAP- and CLIP-tags were introduced into the stator *a*-subunit and rotor $\varepsilon$-subunit, respectively. The specific labeling and the catalytic activity of the purified $F_oF_1$ was confirmed. Furthermore, labeling conditions *in vivo* were examined using benzylguanine-TMR (tetramethylrhodamine) and benzylcytosine-Alexa647 (a Cyanine-5 derivative). So far, efficient labeling was achieved by electroporation into the *E. coli* in stationary phase and localization to the membrane has been observed. Single-cell measurements showed a FRET level of ~0.5 as expected from the structural model of $F_oF_1$.

The later task, lowering copy numbers of labeled $F_oF_1$ in *E. coli*, was tackled by culturing the cells after labeling. After cell divisions for several times, the copy number was expected to decrease to less than ten. Our results revealed that single-molecule imaging of $F_oF_1$ in *E. coli* was actually possible with this procedure.

## 2 FRET-LABELING A SINGLE ENZYME *IN VIVO*

### 2.1 Labeling a SNAP-tag plus a tetracysteine motif on $F_oF_1$-ATP synthase in *E. coli*

The major problem to solve for *in vivo* FRET measurements is the specificity of targeting only the protein of interest. Cysteine or lysine labeling approaches are not suitable due to the abundance of these amino acids. For $F_oF_1$-ATP synthase in living *Escherichia coli*, the fusion of an autofluorescent protein "enhanced green fluorescent protein" (EGFP) to the C-terminus of the non-rotating *a*-subunit has been shown not to interfere with the function of the enzyme[21]. This is surprising due to the size of EGFP (barrel-shaped, 2 nm x 5 nm, with a molecular mass of about 25 kDa) and the short connection of only four amino acids as a linker to the C-terminus. However, the photophysics of EGFP including a low photostability and spectral fluctuations as well as fluorescence lifetime fluctuations[28] depending on the laser power limited its use for the intended *in vivo* FRET measurements. Therefore we replaced the EGFP on the *a*-subunit by a similar sized fusion protein, that is, the SNAP-tag[29] (New England Biolabs). The SNAP-tag can be labeled with a variety of commercially avaliable fluorescent ligands including well-known organic fluorophors like tetramethylrhodamine ("BG-Cell-TMR"). In addition, other fluorescent ligands can by synthesized easily to optimize the Förster radius $R_0$ for 50% energy transfer in the FRET measurement. We synthesized the SNAP ligands BG-Atto565, BG-Atto590 and BG-Atto647N using fluorophores as NHS-esters (Atto-tec).

To monitor subunit rotation in a single $F_oF_1$ in a living bacterium, intramolecular FRET between a rotating and a static subunit is intended. Thus, one of the three subunit of the rotor, i. e. the γ-, ε- or *c*-subunits has to be labeled with the second fluorophore. A small but specific label is a biarsenic dye "FlAsH" (a fluorescein derivative; Invitrogen) which specifically binds to given sequences of 4 cysteines in the order C-C-X-Y-C-C in an α-helix of a protein[30]. FlAsH has been used previously to label $F_oF_1$-ATP synthases in the mitochondria of yeast cells[31, 32]. Procedures to get the dye into the cell and to remove excess unbound fluorophores afterwards have been published[33] and were applied here. A tetracysteine amino acid sequence was fused to the C-terminal-truncated ε-subunit in the $F_1$ part. *In vivo* labeling of the *E. coli* enzyme was facilitated by cold-shock treatment or electroporation. After washing the cells, controlled growth and subsequent washing steps removed unbound FlAsH.

To evaluate the FlAsH fluorescence labeling in *E. coli*, we used a custom-designed confocal microscope[34-40] with a three dimensional piezo scanner to move the bacteria through the laser focus (excitation with 60 ps pulsed excitation at 488 nm with 80 MHz repetition rate; PicoTa490, Picoquant). Time resolved fluorescence images (FLIM) were recorded by two single-photon counting avalanche photodiodes (APDs, SPCM AQR-14, Perkin Elmer) and time-correlated single photon counting electronics (SPC-152, Becker&Hickl) for 4 ms per pixel. Out-of-focus fluorescence was rejected after passing a pinhole. The two spectral ranges in FLIM were 497-568 nm for FlAsH as FRET donor fluorophore, and λ> 595 nm for the two different FRET acceptor dyes Atto590 or Atto647N, respectively.

As shown in the confocal scanning images in Figure 2, FlAsH-labeling of the living *E. coli* cells resulted in homogeneous staining with low fluorescence intensities (Fig. 2A and 2D). The mean lifetime for $F_oF_1$-ATP synthase-bound FlAsH in the absence of a FRET acceptor was $\tau_{FlAsH}$ = 3.4 ns (Fig 2E), with a significant contribution of a second autofluorescence component $\tau$ < 1 ns. For FRET-labeling, the SNAP-tag was modified by either BG-Atto590 or BG-Atto647N, resulting in a shorter lifetime for FlAsH of $\tau_{FlAsH, FRET}$ = 2.55 ns for some *E. coli* cells (Fig. 2B). FRET efficiencies were calculated per pixel using the corrected fluorescence intensities of FRET donor ($I_D$) and acceptor ($I_A$) and the simple formula for the proximity factor P:

$$P = I_A / (I_D + I_A) \qquad (1)$$

Based on the proximity factor images, FRET-labeling of both the tetracysteine on the ε-subunit and the SNAP-tag on the *a*-subunit of the same $F_oF_1$-ATP synthase seemed to be successful (Fig. 2C versus 2F). However, the very low photostability prevented the further use of FlAsH as the donor fluorophore for single-molecule FRET measurements in living *E. coli* cells.

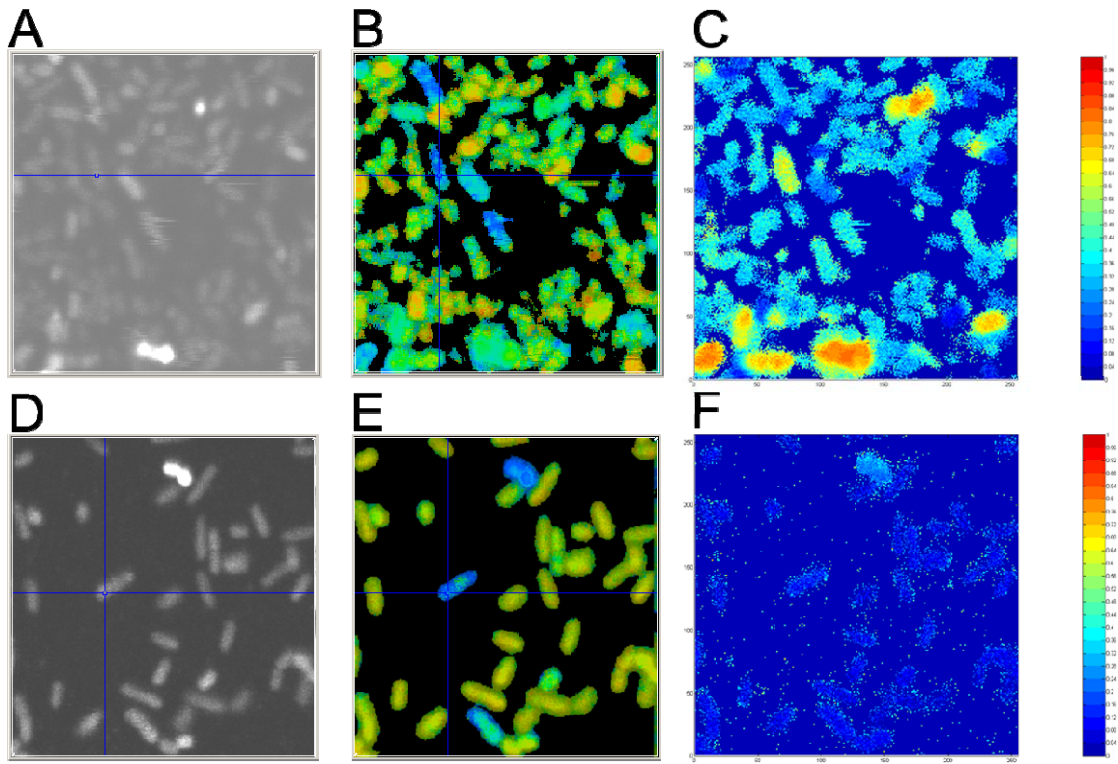

**Figure 2: Confocal fluorescence lifetime image (FLIM) of living *E. coli* cells using $F_oF_1$-ATP synthase labeled with FlAsH and pulsed excitation (60 ps, 80 MHz) at 488 nm.** FRET-FLIM were recorded 2 hours after labeling the cells with FlAsH and BG-Atto647N. Fluorescence intensities of FRET donor FlAsH bound to the tetracysteine motif in subunit ε of $F_oF_1$ (**A, D**). False-colored fluorescence lifetime images (**B, E**) and FRET efficiency images (**C, F**) from the proximity factor $P=I_A/(I_D+I_A)$. Image sizes are 20 x 20 μm. (**A, B, C**) *E. coli* images of $F_oF_1$ in the presence of FRET acceptor Atto647N bound to the SNAP-tag on subunit *a*. One FRET-labeled *E. coli* cell is highlighted by a cursor: (**A**) mean FlAsH intensities are lower due to FRET, (**B**) fluorescence lifetime is about 2.55 ns and is "light blue"-coded [non-labeled younger *E. coli* show lifetimes τ<< 1 ns due to autofluorescence and are "yellow"-coded] and (**C**) the FRET efficiency image indicates that the highlighted cell exhibits an apparent FRET efficiency around 0.6 to 0.7 ("green-yellow" pixels). (**D, E, F**) *E. coli* control images of $F_oF_1$ in the absence of FRET acceptor Atto647N. One *E. coli* cell stained with FlAsH is highlighted by a cursor: (**D**) FlAsH intensities per cell vary, (**E**) fluorescence lifetime is about 3.4 ns at the cursor and is coded "dark blue" [but non-labeled younger *E. coli* show lifetimes τ<< 1 ns due to autofluorescence and are color-coded in "yellow"], and (**F**) the FRET efficiency image indicates that the highlighted cell exhibits no FRET efficiency and, accordingly, apparent FRET is less than 0.2 (which is due to spectral cross talk). The color bars to the right correspond to panels C and F.

## 2.2 Labeling a SNAP- plus a CLIP-tag on purified $F_oF_1$-ATP synthase

Following the first approach of labeling the rotating ε-subunit *via* FlAsH, we developed the fusion of the CLIP-tag to the truncated ε-subunit[41]. The CLIP-tag is a 20 kDa mutant protein of the $O^6$-alkylguanine-DNA alkyltransferase and is labeled by a very similar strategy like the SNAP-tag. The $NH_2$-modified benzylcytosine ligand for the CLIP-tag can be functionalized by any NHS-ester of rhodamines, oxazines, cyanines and others. Combining SNAP- and CLIP-tags in one $F_oF_1$ allows for fine-tuning the FRET parameters and the photon counting efficiencies for any given laser system or detection apparatus. The status of the SNAP/CLIP FRET-labeled $F_oF_1$ is summarized in Fig. 3. Specific labeling was achieved *in vitro* for the SNAP-tag with either BG-Alexa488, BG-Cell-TMR or BG-Alexa647, and for the CLIP-tag with BC-Cell-TMR.

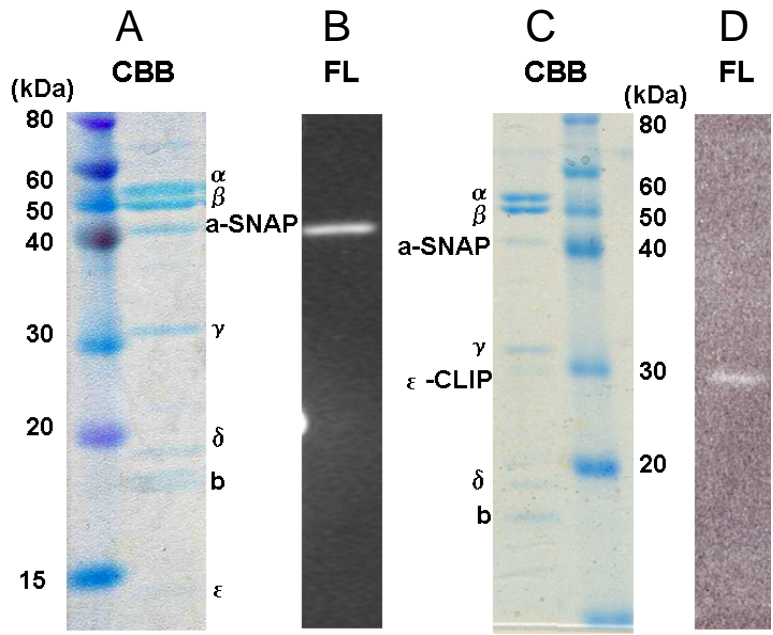

**Figure 3: SDS-PAGE of $F_oF_1$-ATP synthase with SNAP-tag fused to the C-terminus of subunit *a* and CLIP-tag fused to C-terminal-truncated subunit ε.** The two lanes on the left (**A**, **B**) show the subunits of the mutant enzyme comprising only a SNAP fusion to the *a*-subunit. The two lanes on the right (**C**, **D**) show the subunits of the double mutant enzyme comprising the CLIP fusion to the truncated ε-subunit and the SNAP fusion to the *a*-subunit. (**A**) Coomassie Blue (CBB) stained subunits of $F_oF_1$-ATP synthase showing *a*-SNAP and the protein standard with molecular weights in kDa on the left part. (**B**) Fluorogram (FL) of the same preparation after staining the SNAP-tag with BG-Cell-TMR. (**C**) Coomassie Blue stained subunits of $F_oF_1$-ATP synthase showing ε-CLIP and *a*-SNAP with molecular weight standards on the right part. (**D**) Fluorogram of the same preparation after staining the CLIP-tag with BC-Cell-TMR.

Afterwards, the FRET-labeled $F_oF_1$-ATP synthase was reconstituted as a single enzyme per liposome, and single-molecule FRET measurements were accomplished using the freely diffusion enzymes in a droplet of buffer solution as described previously. We modified our confocal microscope to use continuous laser excitation at 532 nm (COMPASS 315M, Coherent) for the FRET donor TMR, in combination with a single 60 ps pulse at 635 nm within 50 ns (LDH-P-635B, Picoquant) for probing the existence of the FRET acceptor dye simultaneously. A dual-line notch filter 555/646 redirected the laser beams in epi-fluorescence configuration. They were focussed into the solution by a water immersion objective (UPlanSApo 60×W, 1.2 N.A., Olympus). Fluorescence was detected by two APDs simultaneously (AQR-14, Perkin Elmer) in two spectral ranges separated by a dichroic beam splitter (DCXR640, AHF Tübingen). Briefly, fluorescence passed a 150 μm pinhole. Fluorescence of TMR was detected between 545 nm and 625 nm (HQ 585/80, AHF) and the signal from Alexa647, the FRET acceptor, between 663 nm and 737 nm (HQ 700/70, AHF).

Fluorescence photons were sorted with respect to the red laser pulse to generate two parallel time trajectories using our custom-made software 'Burst_Analyzer' by N. Zarrabi (University of Stuttgart). FRET-labeled $F_oF_1$-ATP synthases were identified by photon bursts which exceeded the minimum count rate threshold for the FRET acceptor test. The mean burst duration was about 30 ms as expected for freely diffusing liposomes without trapping[42], but also bursts with several hundred millisecond durations were found. FRET efficiency ($E_{FRET}$) levels within these bursts were assigned manually. As shown in Fig. 4A, ATP-driven rotation of the ε-subunit (stopping positions as light stars) with respect to the *a*-subunit (symbolized by a dark star) of $F_oF_1$-ATP synthase should result in a well-defined sequence of three distinct FRET efficiencies. Assigned FRET levels in $F_oF_1$-ATP synthase showing three or more FRET levels in a photon burst were added into histograms. In the FRET efficiency histogram in Fig. 4B, the low FRET position of the

CLIP-tag on ε is located around $E_{FRET}$= 0.3, the medium around $E_{FRET}$= 0.5, and the high around $E_{FRET}$= 0.75. The levels were nearly equally distributed, and the dwell times were found in the range of 12 ± 2 ms. The preferred FRET level sequence for ATP hydrolysis was L→M→H→, but most of the bursts showed alternating or oscillating FRET level changes.

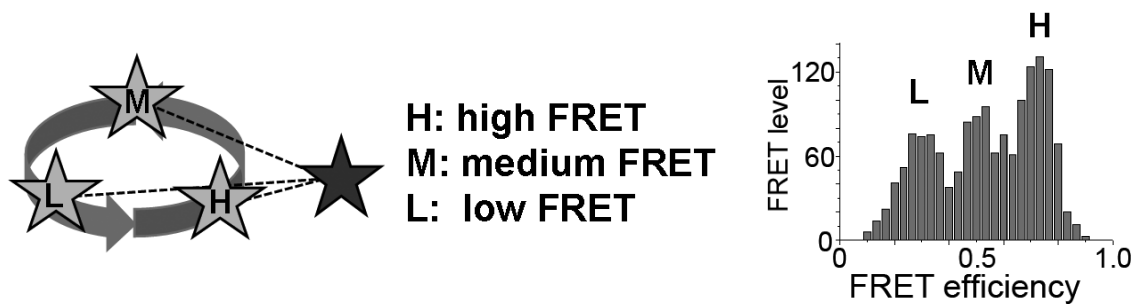

Figure 4: Three distinct FRET efficiencies of $F_oF_1$-ATP synthase with TMR-labeled SNAP-tag on subunit *a* and Alexa647-labeled CLIP-tag on subunit ε. Left, model of expected distances and FRET efficiencies during stepwise rotation of the ε-subunit with respect to the *a*-subunit. Right, experimental FRET efficiency distribution in the presence of ATP measured by single-molecule FRET with enzymes reconstituted into liposomes.

## 3 SINGLING OUT ONE ENZYME IN VIVO

### 3.1 Labeling a SNAP- plus a CLIP-tag on $F_oF_1$-ATP synthase in living *E. coli*

We tried labeling of SNAP- and CLIP-tags of $F_oF_1$ expressed in living *E. coli*. The FRET donor dyes (BG-Cell-TMR for the SNAP-tag or BC-Cell-TMR for the CLIP-tag) were permeable to the cell membrane, and $F_oF_1$ in the inner membrane of *E. coli* could be labelled by adding the dye (final concentration was 1 μM) to the cell suspended in LB medium (O.D.$_{600}$ ~ 2) and incubating for 1-2 hrs at 37°C. However, the FRET acceptor dye BG-Alexa647 was not permeable to the membrane and incapable of reacting with $F_oF_1$ *in vivo* under the condition described above. So, introduction of donor and acceptor dyes into the cytoplasm of the cell was carried out using electroporation. Cells grown in 3 ml of LB medium (12-16 hrs) to the stationary phase were centrifuged at 7000 rpm for 5 min at room temperature, and suspended with 10% glycerol. Centrifugation and suspension were repeated three times, and finally suspended with 500 μl buffer containing 10% glycerol. BG-Alexa647 and BC-Cell-TMR were added to 100 μl of the cell suspension, incubated 5 min on ice, and applied to the BioRad Micro Pulser (1.8 kV, 1 pulse). Then, 1 ml of ice-cold SOB medium was added, and incubated for 1-2 hrs at 37°C. After incubation, cells were centrifuged at 3000 rpm for 5 min at room temperature and resuspended with 1 ml of M9 medium. Centrifugation and suspension were repeated three times, and finally suspended with 100 μl of M9 medium and used for observation.

Cells labeled by BG-Alexa647 (acceptor) and BC-Cell-TMR (donor) were observed under an objective-type total internal reflection fluorescence (TIRF) microscope[43, 44] constructed on an inverted microscope (IX71, Olympus) equipped with a 60× oil immersion objective lens (PLAPON60XOTIRFM, NA=1.45, Olympus). The donor was selectively excited with continuous-wave laser light at 532 nm (DPGL-2100F, Photop). Fluorescence from donor and acceptor was separated by a dichroic mirror (Chroma 630DCXR) and further selected by band-pass filters (Chroma ET585/65M for donor, Semrock FF01-692/40 for acceptor) equipped in a dual-view unit (Optical Insights)[45, 46]. Images of FRET donor and acceptor were projected onto a CCD camera (2400-75AH, Hamamatsu Photonics) coupled with an image intensifier (C8600, Hamamatsu Photonics) and recorded with a digital video recorder (DSR-11, Sony) at 30 frames/s. Fig, 5 shows a typical donor (D) and acceptor (A) images of single *E. coli* cells. On average, FRET efficiency of 0.51 (N=68) was obtained at the single-cell level (Fig. 5, top). This ensemble value was consistent with FRET efficiencies obtained by the single-molecule FRET *in vitro* (Fig. 4). To confirm FRET, the acceptor was selectively photobleached by an intense excitation with a continuous-wave laser light at 633 nm (5-LHP-151, Melles Griot) (Fig. 5, middle). After phobleaching of the acceptor, fluorescence intensity of the donor obtained by excitation at 532 nm increased, indicating that FRET did occur in the cell (Fig. 5, bottom).

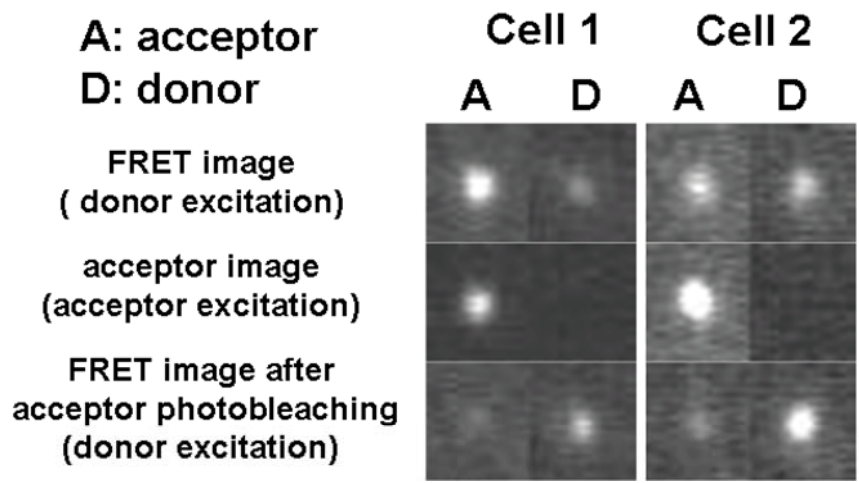

**Figure 5: FRET in single TMR(donor)- and Alexa647(acceptor)-labeled $F_oF_1$-ATP synthases in living *E. coli* cells.** (**Top**) Donor and acceptor images of single cells obtained by excitation at 532 nm. (**Middle**) Images obtained by excitation at 633 nm. (**Bottom**) Images obtained by excitation at 532 nm, after selective photobleaching of the acceptor.

### 3.2 Dilution of TMR-labeled $F_oF_1$ *in vivo* by culturing cells after labeling

The steady state expression levels of the β-subunit of $F_oF_1$ encoded in the chromosome of single *E. coli* cells were reported to be in the range of several hundred copies[47]. In accordance with this report, the density of the TMR-labeled SNAP-*a* $F_oF_1$ in *E. coli* was so high that individual molecules could not be resolved as single diffraction-limited spots in the fluorescence images. In the cell, various proteins are continuously synthesized and degraded. The lifetime (i.e. the duration before degradation after synthesis) of a specific protein is determined by the "N-end rule"[48]. According to this rule, the lifetimes of $F_oF_1$ are expected to be much longer than the doubling time of the growing *E. coli*. In the process of cell growth and division, unlabeled $F_oF_1$ will be newly synthesized and supplied, and the fraction of TMR-labeled $F_oF_1$ in each cell will decrease. Taking advantage of these properties, we tried to decrease the copy number of TMR-labeled $F_oF_1$ in the cell by multiple divisions.

The TMR-labeled cells were cultured in M9 medium with 0.45% glucose at 32°C and observed under an inverted microscope (IX71, Olympus) equipped with a stage heater (Tokai Hit). For phase-contrast and epi-fluorescence images we used a 100× oil immersion objective lens (UPLFLN100XO2PH, Olympus), a mirror unit (U-MWIG3, Olympus) and a CCD camera (WAT-120N, Watec). During incubation, *E. coli* showed multiple cell divisions (Fig. 6, left). The epi-fluorescence images of TMR revealed that fluorescence intensity of the cells decreased and became hardly visible after multiple cell divisions. When the labeled cells were cultured in M9 medium with 0.45% glucose in a test tube at 37°C overnight and imaged under the TIRF microscope described above, individual TMR-labeled $F_oF_1$ molecules were observed as slowly diffusing spots on the cell membrane (Fig. 6, right). These results clearly indicate that the copy number of TMR-labeled $F_oF_1$ can be decreased down to several molecules in single cells, making it possible to conduct single-molecule FRET measurement in living cells.

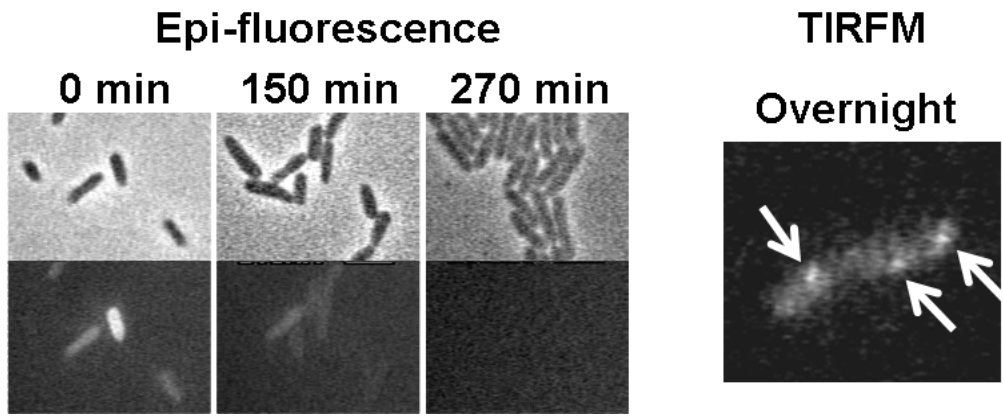

**Figure 6:** Single-molecule imaging of TMR-labeled $F_oF_1$-ATP synthases by culturing cells after labeling. (**Left**) Phase-contrast (**Top**) and epi-fluorescence image (**Bottom**) of *E. coli* cells showing multiple consecutive divisions. The mean brightness decreased with cell divisions. (**Right**) Fluorescence image showing single TMR-labeled $F_oF_1$ molecules (arrows) in an *E. coli* after overnight cell culture. The image was obtained by a TIRF microscope.

## 4 CONCLUSIONS

Genetically encoded enzyme- and peptide-based tags capable of specific labeling with fluorescent dyes in living cells were successfully utilized to introduce FRET donor and acceptor fluorophores into rotor and stator subunits of the rotary motor protein $F_oF_1$-ATP synthase. In single-molecule measurements *in vitro*, $F_oF_1$ labeled by donor and acceptor showed three distinct FRET levels corresponding to the three pausing positions of the ε-subunit of $F_oF_1$ during rotation. FRET was also detected at the single-cell level after direct labeling of $F_oF_1$ expressed in living *E. coli*. Real-time single-molecule imaging of $F_oF_1$ in living *E. coli* was possible after multiple divisions of the labeled cells. With this strategy, we are now trying to measure single-molecule FRET of $F_oF_1$ *in vivo*. Our approach can be applied not only to $F_oF_1$ but also other proteins, and will be a powerful method to unravel the conformational dynamics of various biological nanomachines working in living cells.

**Acknowledgements**

This work was partly supported by Grant-in-Aid for Scientific Research on Priority Areas (Project ID 22018018 to R.I.) from the Ministry of Education, Culture, Sports, Science and Technology, Japan, and by Japan-Germany Bilateral Joint Projects from Japan Society for the Promotion of Science (to R.I.). Financial support from the German Science Foundation (DFG project BO1891/10-2 to M.B.) is gratefully acknowledged.